\documentclass[11pt]{emulateapj}\usepackage{times}

\usepackage[usenames]{color}

\usepackage{graphicx}
\usepackage{epstopdf}
\usepackage{url}

\newcommand{\be} {\begin{equation}}

\newcommand{\ee} {\end{equation}}

\newcommand{\xmm}{{\em XMM--Newton}}
\newcommand{\XMM}{{\em XMM--Newton}}

\newcommand{\CXO}{{\em Chandra}}
\newcommand{\bc}{\begin{center}}
\newcommand{\ec}{\end{center}}

\def\ltsima{$\; \buildrel < \over \sim \;$}
\def\lsim{\lower.5ex\hbox{\ltsima}}
\def\loe{\lower.5ex\hbox{\ltsima}}
\def\gtsima{$\; \buildrel > \over \sim \;$}
\def\gsim{\lower.5ex\hbox{\gtsima}}
\def\goe{\lower.5ex\hbox{\gtsima}}

\def \cm2{cm$^{-2}$\,}

\def\ergs {erg\,s$^{-1}$}
\def\ergscm2 {erg\,s$^{-1}$cm$^{-2}$}

\def\srca{LS\,I\,+61$^{\circ}$303}
\def\srcb{LS\,5039}
\def\srcc{HESS\,J0632+057}

\begin{document}

\shorttitle{\textsc{The TeV binary \srcc\ in the low and high X-ray state}}
\shortauthors{\textsc{Rea \& Torres}}

\title{\textsc{The TeV binary \srcc\ in the low and high X-ray state}}

\author{Nanda Rea\altaffilmark{1} \& Diego F. Torres\altaffilmark{1,2} %
}

\altaffiltext{1}{Institut de Ci\`encies de l'Espai (IEEC-CSIC),
              Campus UAB,  Torre C5, 2a planta,
              08193 Barcelona, Spain}
\altaffiltext{2}{Instituci\'o Catalana de Recerca i Estudis Avan\c{c}ats (ICREA).}

\begin{abstract}

We report on a 40\,ks \CXO\ observation of the TeV emitting high
mass X--ray binary \srcc\ performed in February 2011 during a high-state of X-ray and TeV
activity.  We have used the ACIS-S camera in Continuos Clocking mode
to search for a possible X-ray pulsar in this system. Furthermore, we compare the emission of the source during this high state, with its X-ray properties during a low state of emission, caught by a 47\,ks \XMM\ observation on September 2007. We did not find
any periodic or quasi-periodic signal in any of the two observations.  We derived an average pulsed fraction 3$\sigma$ upper limit for
the presence of a periodic signal of $\lsim$35\% and 25\% during the low and high emission state, respectively (although this limit
is strongly dependent on the frequency and the energy band). Using the best X-ray spectra derived to date for \srcc, we found evidence for a significant spectral change between the low and high X-ray emission states, with the absorption value and the photon index varying between $N_{\rm H} \simeq 2.1-4.3\times10^{21}$\cm2\,and $\Gamma \simeq 1.18-1.61$. At variance with what observed in other TeV binaries, it seems that in this source the higher the flux the softer the X-ray spectrum.

\end{abstract}


\section{Introduction}

X-ray binary systems are one of the few astronomical objects that, under some
conditions, are expected to appear as point-like TeV emitting sources when observed by instruments having 
the current sensitivity. That is the case of the three
well-established members of the class, PSR\,B1259--63 (Aharonian et
al.~2005a), \srca\ (Albert et al.~2006), and \srcb\ (Aharonian et
al.~2005b,~2006). Among these three systems, PSR\,B1259--63 is composed by a 48\,ms pulsar in a 3.4\,ys orbit with a Be companion, \srca\, and \srcb\, have a $\sim$26 and 4\,days orbital periods, and host a Be and an O start, respectively.  Unfortunately the nature of the compact objects in these two binary systems is still unkown. 

Of the many tens of unidentified sources discovered in the Galactic Plane H.E.S.S. survey (the survey done with
the High Energy Stereoscopic System), \srcc\ is
one of only two unidentified very-high-energy gamma-ray sources which
appear to be point-like within the experimental resolution (the other is
coincident with the gravitational centre of the Milky Way; Aharonian et al. 2007).

Follow-up observations of \srcc\ with \XMM\ have revealed an X-ray
source (XMMU J063259.3+054801) coincident with the TeV detection as
well as with the massive star MWC 148 (spectral type B0pe), at a distance of $\sim1.5$\,kpc (Hinton et
al. 2009). The chance coincidence of a massive star within the 1
arcsec error box of the brightest X-ray source in the
\XMM\ observation has been quantified by the latter authors to be of
the order of $10^{-6}$.  XMMU J063259.3+054801 has in addition been
found to have a hard non-thermal X-ray spectrum, and significant variability on hour
timescales.  These features are similar to those found in the many
X-ray observations of \srca\ and \srcb\ (Sidoli et al. 2006, Esposito et al. 2007; Kishishita et al. 2009; Rea et al. 2010, 2011 and
references therein), strengthening the association between the TeV source and XMMU J063259.3+054801/MCW 148, and hence establishing the possibility for \srcc\ to be the fourth TeV binary.

Observations conducted with the Giant Metrewave Radio Telescope (GMRT)
and the Very Large Array (VLA) revealed the radio counterpart of \srcc: a point-like, low-flux, variable radio
source at the position of MCW 148 was detected in both 1280 MHz with
GMRT and 5 GHz with VLA (non simultaneously), with an average spectral index of --0.6 which
can be generated by synchrotron-emitting electrons (Skilton et
al. 2009).

Subsequent observations of \srcc\ with the VERITAS array detected no
significant signal from it, excluding that the source is a steady
gamma-ray emitter (Acciari et al. 2009, Maier et
al. 2009). Simultaneous to these TeV observations, an X-ray campaign
conducted with \emph{Swift}-XRT revealed a significant variability in the X-ray emission (Falcone et
al. 2010).  From this early X-ray monitoring there was no signature of an orbital periodicity for \srcc, however, assuming that the spectral variability is due to an orbital modulation, the
orbital period was estimated to be larger than $\sim 54$ days. A similar
conclusion was reached by Aragona et al. (2010) through optical spectroscopy of the massive star.  Confirming the
previous constraints, Bongiorno et al. (2011) recently 
reported an orbital period of  320$\pm$5 days, using data from the continuing long-term \emph{Swift}-XRT
monitoring campaign.  By analogy with \srca\ and \srcb, the compact object
hosted in \srcc\ might then be a young non-accreting pulsar,
or an accreting compact object (black hole or neutron star) driving a jet.

\begin{figure*}
\includegraphics[height=7cm,width=9cm]{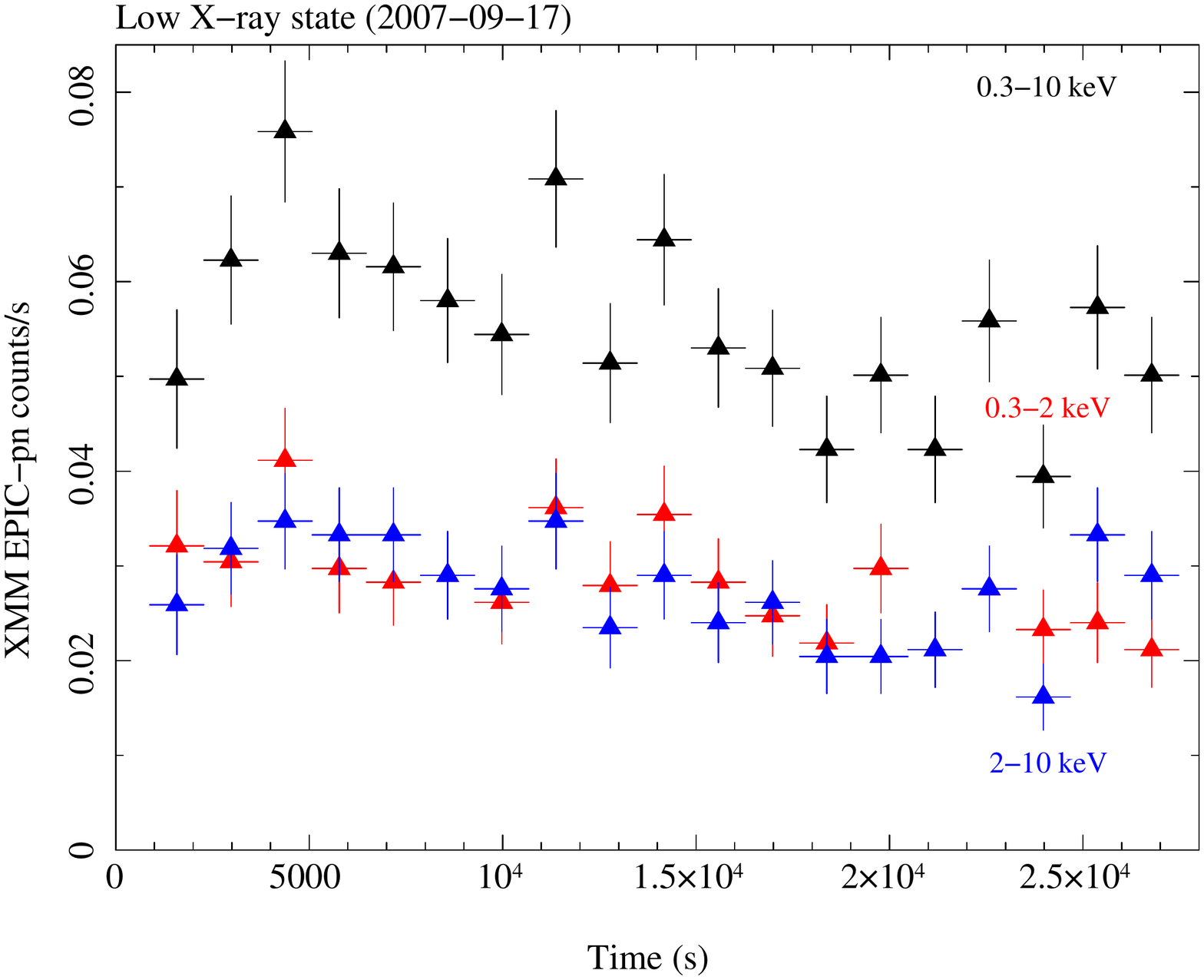}
\hspace{-1.cm}
\includegraphics[height=7cm,width=9cm]{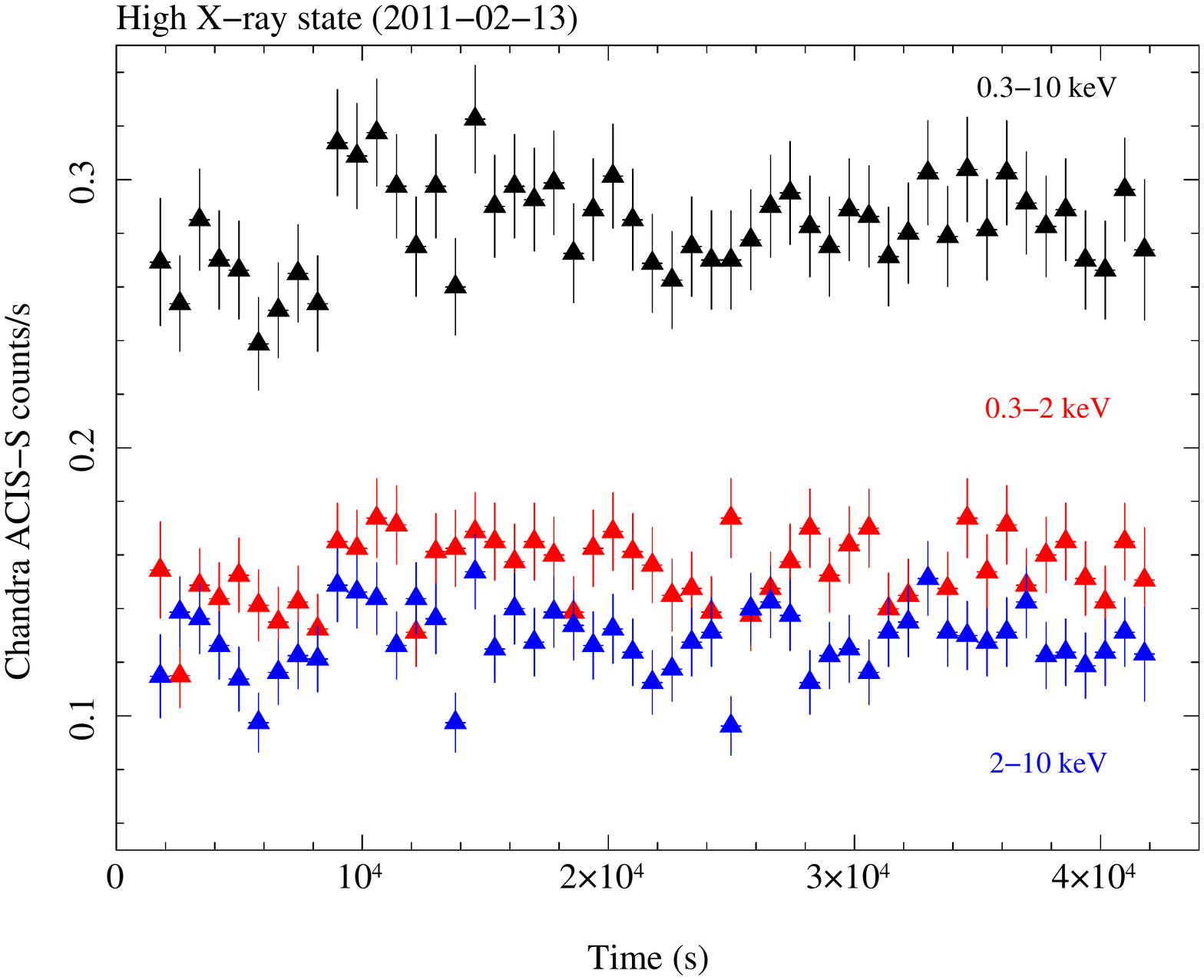}
\caption{\XMM\, (left panel) and \CXO\, (right panel) back-ground subtracted light-curves in the 0.3--10, 0.3--2 and 2--10 keV band, binned at 1400\,s and 700\,s, respectively.}
\label{lc}
\end{figure*}

On 2011 January 23rd, \emph{Swift}-XRT detected a rise in the
X-ray flux of \srcc\,(Falcone et al. 2011). The flux increase was a factor of $\sim 3$, which
appeared similar to the rises that occurred $\sim 320$ and $\sim 640$
days before. We now know that this is the orbital period of the binary
(Bongiorno et al. 2011).
Motivated by this increase in X-ray activity, TeV observations
were conducted by VERITAS on 2011 February 7-8, which
detected the source at higher TeV flux than during the previous VERITAS campaigns
(Ong et al. 2011). These results were soon confirmed by the MAGIC
collaboration (Mariotti et al. 2011).  Contemporaneous with the X-ray
increase, further radio observations were also conducted, which
suggested the presence of slightly extended radio emission at
milli-arcsec scales coincident with the position of the Be companion (Moldon et al. 2011).
However, simultaneous optical observations conducted from 2011 January 5 and February 24,  revealed no significant change in the radial velocities of the Be companion star (Casares et al. 2011).

During this high TeV/X-ray emission period we have conducted a 40\,ks
observation with the \CXO\ X-ray Observatory kindly granted to us
using the Discretionary Director Time (DDT). Here we report on the
spectral and timing characteristics of \srcc\ during this high X-ray emission state,
and we compare them with an archival \XMM\ observation (Hinton et
al. 2009) performed during a low X-ray emission state.

\begin{figure*}
\vbox{
\hbox{
\includegraphics[height=7cm,width=9cm]{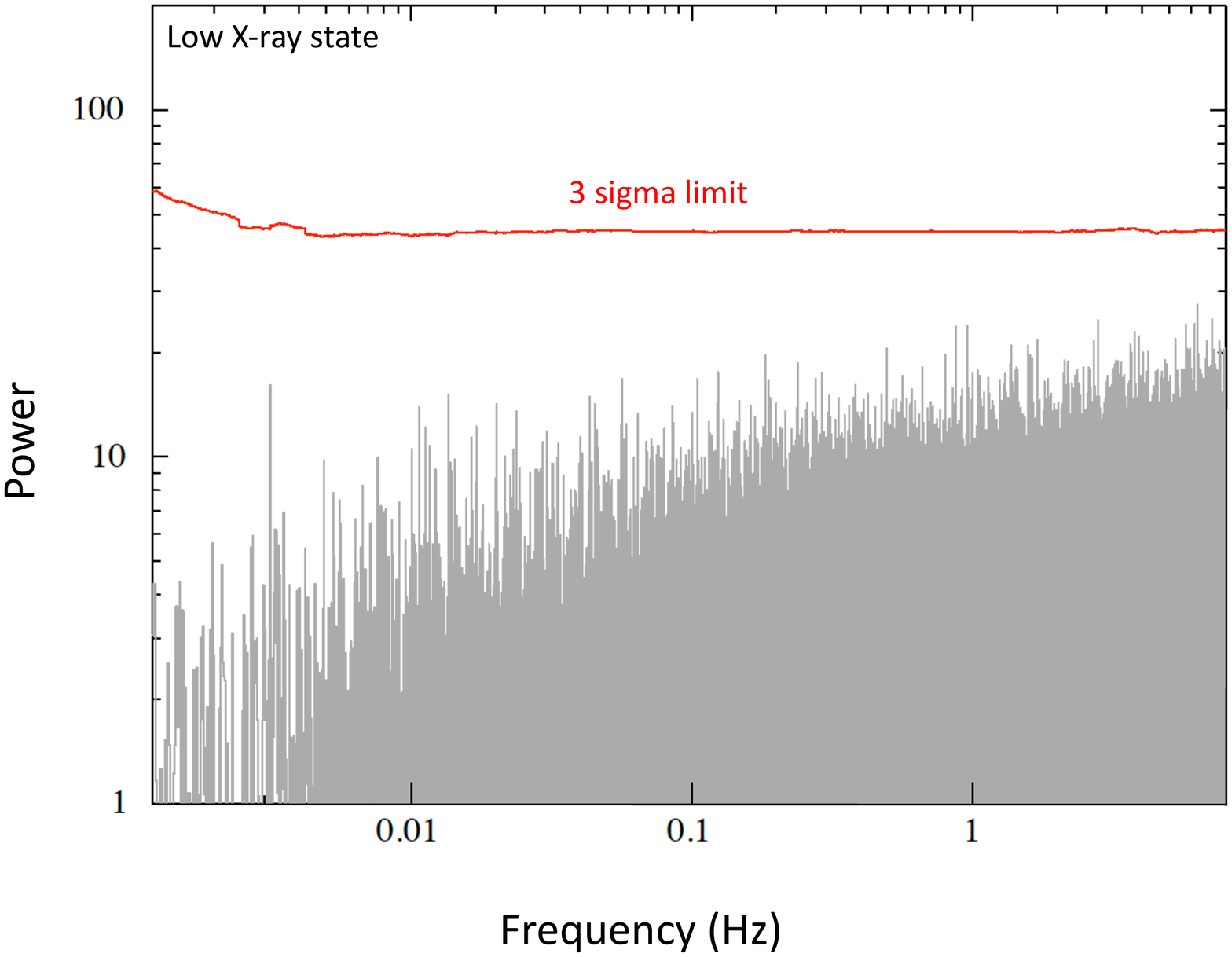}
\hspace{-1.4cm}
\includegraphics[height=7cm,width=9cm]{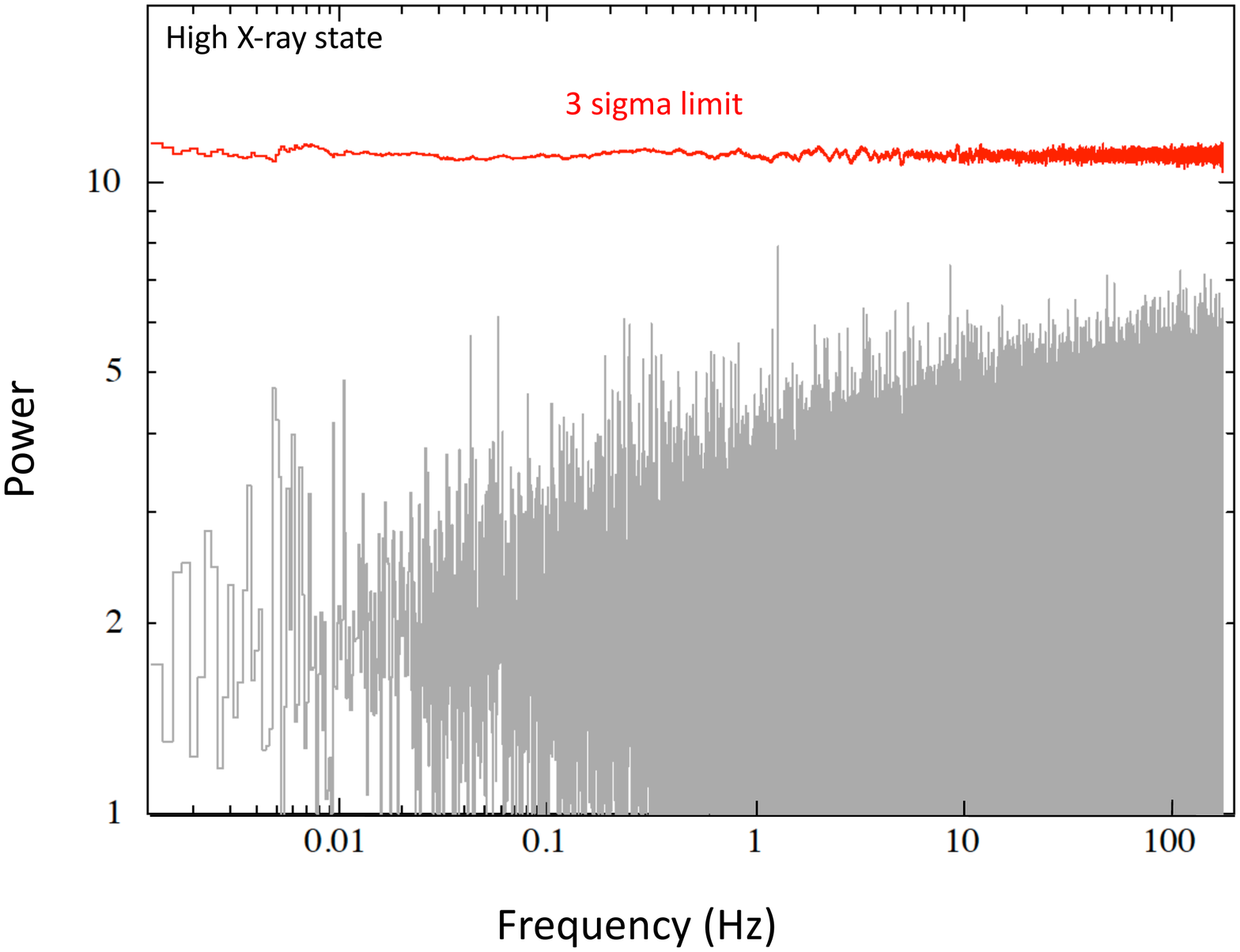}}
\vspace{-0.6cm}
\hbox{
\includegraphics[height=7cm,width=9cm]{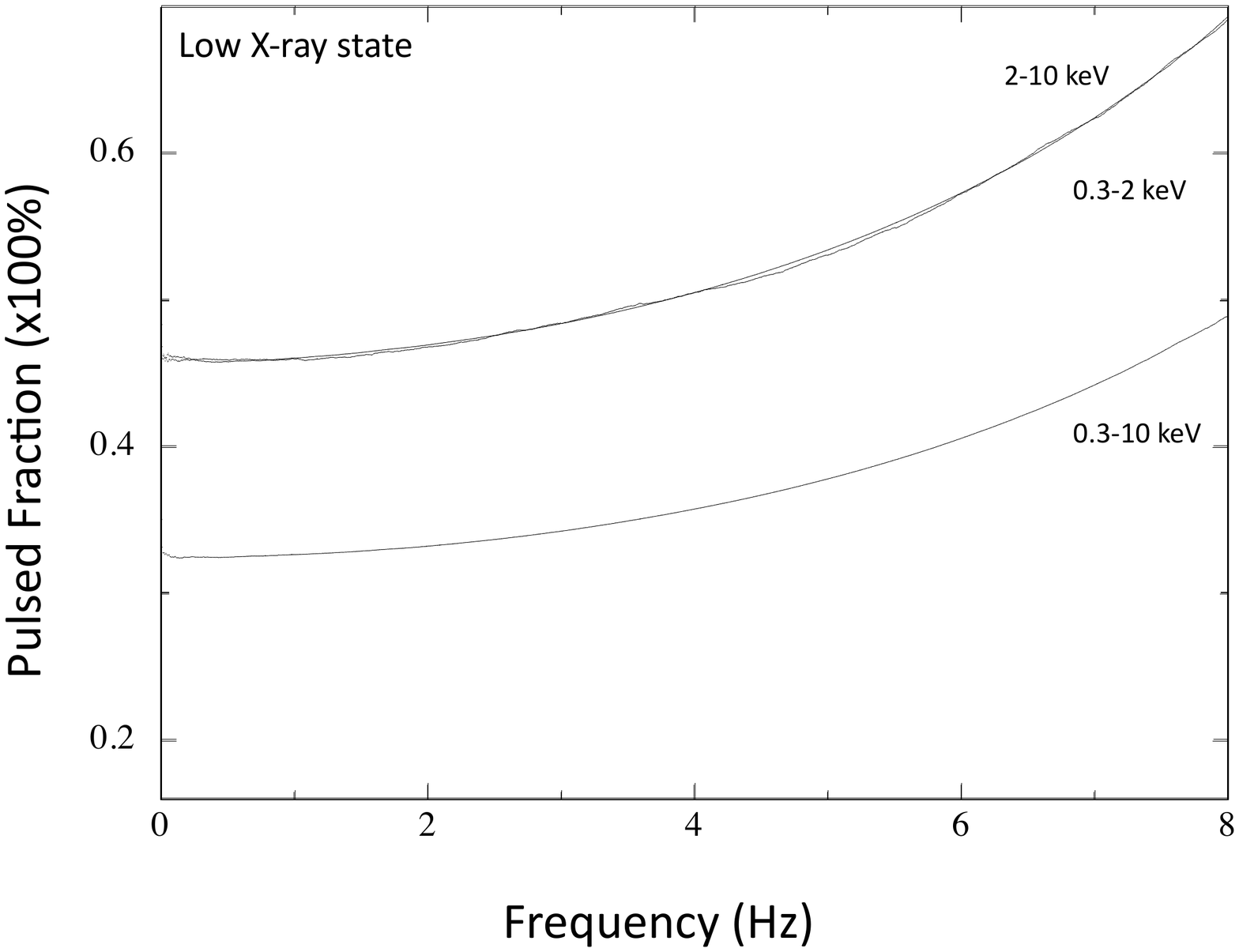}
\hspace{-1.1cm}
\includegraphics[height=7cm,width=9cm]{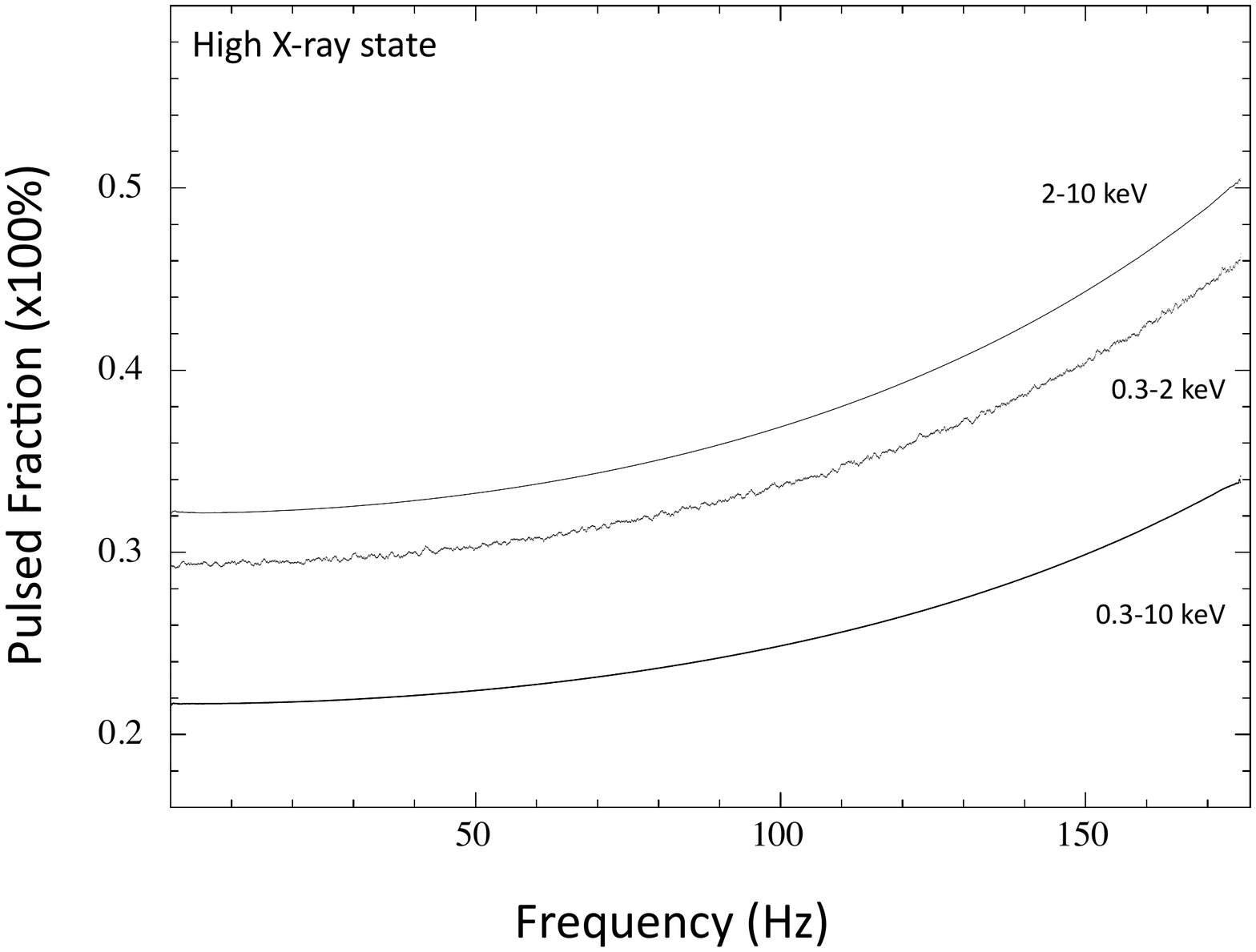}}}
\caption{{\em Left column}:  Timing analysis of the \XMM\, data taken during a low intensity state. {\em Right column}: same as the left column but for the new \CXO\, data taken during a high intensity state of the source. {\em Top panels}:  Power spectra of the two observations with the relative 3$\sigma$ coherent signal detection limit. {\em Bottom panels}:  3$\sigma$ limits on the pulsed fraction of a detectable signal in different energy bands.}
\label{dps}
\end{figure*}

\section{Observations and Data analysis}
\label{data}

The Advanced CCD Imaging Spectrometer (ACIS) camera on board of the \CXO\, observatory (Weisskopf et al. 2000)
observed \srcc, on 2011--02--13 (start time 21:15:13 (UT); Obs-ID
13237) for an exposure time of 40\,ks in Continuos Clocking (CC) mode
(FAINT). We have chosen this mode since it provides a time resolution
of 2.85\,ms, suitable for searching for fast pulsations.  The CC mode also
provides imaging along a single direction. The data analysis mimics that performed in Rea et al. (2010,
2011). Data were reprocessed using the CIAO
software (ver. 4.3) and the \CXO\, calibration files (CALDB
ver. 4.4.3).  The source was positioned in the back-illuminated
ACIS-S3 CCD at the nominal target position (RA 06:32:59.30; Dec
+05:48:01.00; Hinton et al. 2009).  Standard processing of the data was performed by the
{\em Chandra X-ray Center} to level 1 and level 2.

We corrected the times for the variable delay due to the spacecraft
dithering and telescope flexure, starting from level 1 data and
assuming that all photons were originally detected at the target
position. We excluded hot pixels, bad columns, and possible afterglow
events. Finally, photon arrival times are in TDB and were referred to
the barycenter of the Solar System using the JPL-DE405 ephemeris.

Events in the 0.3--10\,keV energy range were extracted from a small region of
5$\times$5 pixels around the source position for timing
analysis, so as to reduce the background contamination in the timing analysis. The source
spectrum was instead extracted from a rectangular region of 5$\times$25 pixels
around the source position, with the background being taken
independently from a source-free region in the same chip.

Response matrix (RMFs) and ancillary response (ARFs) files were
produced first creating a weighted image, re-binning by a factor of
8. We used this re-binned image to build the RMF file using the {\tt
  mkacisrmf} tool, with an energy grid ranging from 0.3 to 10 keV in 5
eV increments. Using this RMF and the aspect histogram created with
the aspect solution for this observation ({\tt asphist}), we generated
the appropriate ARF file for the source position. The source ACIS-S
count rate in the 0.3--10\,keV energy band was $0.265\pm
0.004$\,counts\,s$^{-1}$ (all errors in the text are reported at 90\%
confidence level, unless otherwise specified).  


We have also re-analized the \XMM\, observation reported in Hinton et
al. (2009), in search for X-ray pulsations, and with the aim of comparing the low and
high X-ray state of \srcc\ using the two best available spectra. The
\xmm\, Observatory (Jansen et al. 2001) observed \srcc\, on
2007--09--17 (start time 01:22:50 (UT); Obs-ID 0505200101) for an
exposure time of 47\,ks with the EPIC-pn in Prime Full Extended Windowed mode. The addition of the two MOS cameras gave consistent results, we then decided to use only the pn camera. This observing mode results in a
timing resolution of 199.2\,ms, a 27.2 $\times$ 26.2 arcmin$^2$ field
of view and only 2.3\% of out of time events. Data have been processed
using SAS version 11.0.0 with the most updated calibration files (CCF)
available at the time the reduction was performed (April
2011). Standard data screening criteria were applied in the extraction
of scientific products. 

The last part of the observation was affected
by proton flares which we have removed in our final products, except
for the event file used for the pulsation search (which is not
affected by such kind of flares), to use as much source photons as
possible. Cleaning the events from the proton flare results in a good
exposure time for spectral analysis of 27\,ks. 

We have extracted the
source photons from a circular region of 15$^{\prime\prime}$ (such to
avoid any chip gaps), centered on the source point spread function
(PSF), and the background from a larger region far from the source but
in the same CCD (we have re-scaled the background spectrum to
take into account the different extraction region with respect to the
source one). For the spectral analysis we used only photons with PATTERN$\leq$4, and FLAG=0.
The source EPIC-pn count rate in the 0.3--10\,keV energy
band was $0.064\pm 0.002$\,counts\,s$^{-1}$.

We report in Fig.\,\ref{lc} on the light-curves of the two observations we study here, which clearly show also the short-term X-ray variability of \srcc\, (see also Hinton et al. 2009; Acciari et al. 2009; Falcone et al. 2010; Bongiorno et al. 2011)

\begin{center}
\begin{figure*}
\hbox{
\includegraphics[height=8cm,width=8cm]{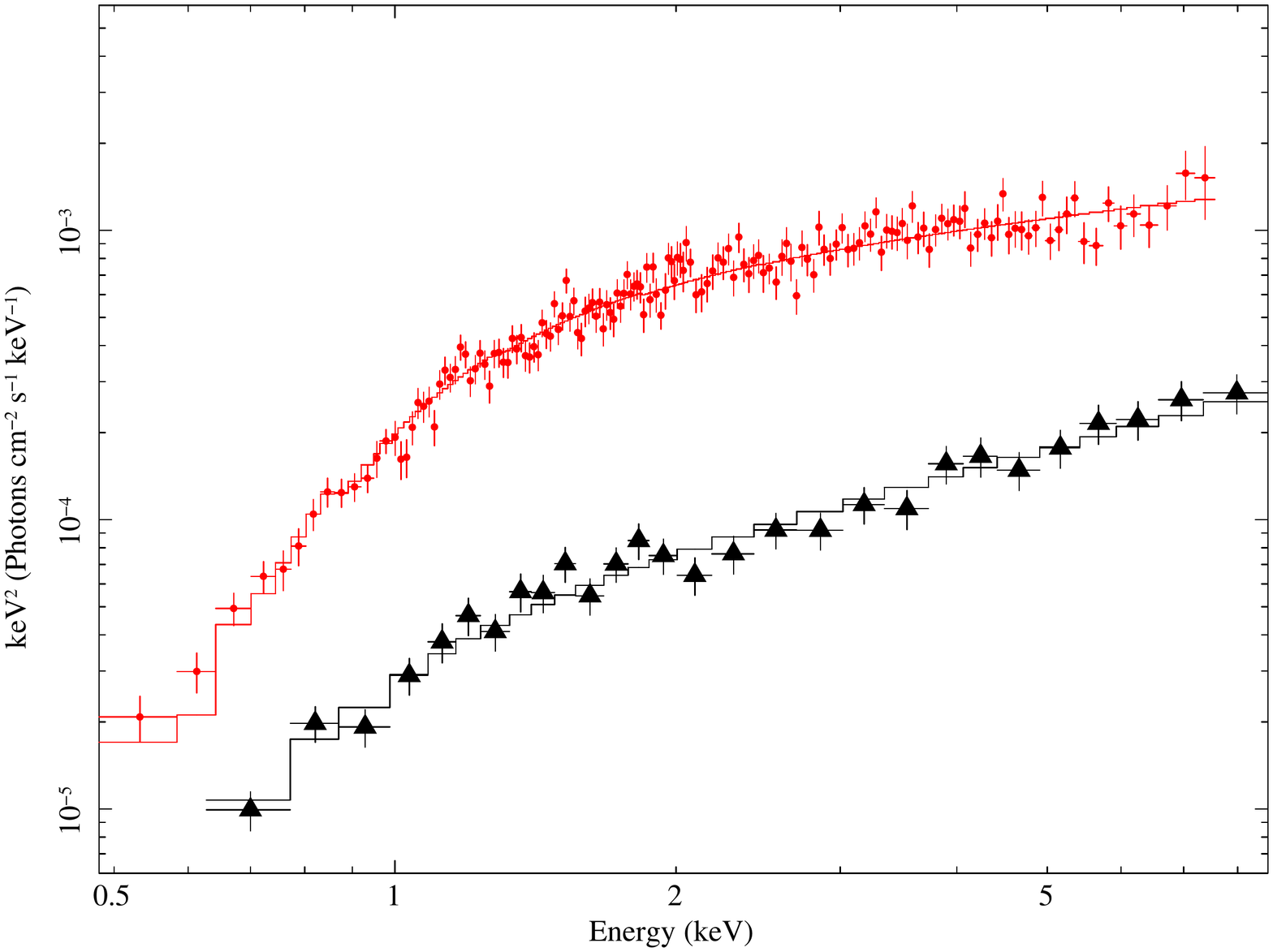}
\includegraphics[height=7cm,width=10cm]{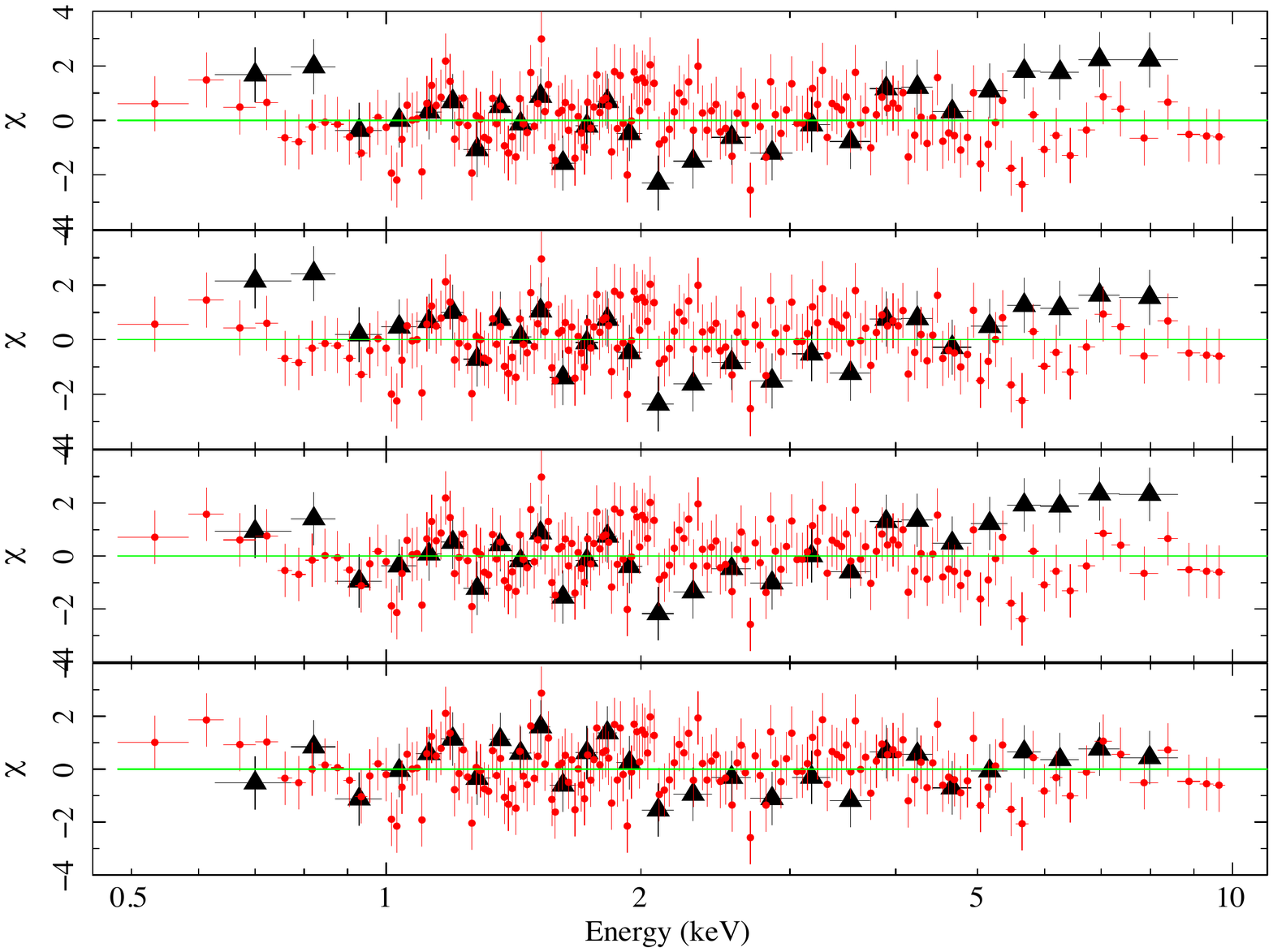}}
\caption{{\em Left panel}: Spectral energy distribution of \srcc\ during the high (red; \CXO) and the low states (black; \XMM) for an absorbed power-law model with all parameters free to vary among the two spectra. {\em Right panel}: residuals of the fits of an absorbed power-law with (from top to bottom) $N_{\rm H}$ and $\Gamma$ equal for both spectra, $N_{\rm H}$ equal and $\Gamma$ free,  $N_{\rm H}$ free and $\Gamma$ equal, and all parameters free (see text for details).}
\label{spectra}
\end{figure*}
\end{center}

\section{Results}
\label{results}

\subsection{Timing analysis: search for pulsations}
\label{timing}

We searched for periodic and quasi periodic signals in the \CXO\ and
\XMM\ observations performing a series of Fast Fourier
Transforms (FFTs; van der Klis 1989). Given the length of our two observations ($\sim$40\,ks
each), the timing resolution of the \CXO\ (2.85\,ms) and \XMM\
(199.2\,ms) observations, and the number of counts of our
observations, we could search for periodic signals in the $0.005
-175$\,Hz and $0.005 -8$\,Hz frequency range,
respectively\footnote{Note that in the search we have oversampled the
  \XMM\ timing resolution by a factor of 3 to reach the 8\,Hz upper
  bound (see Israel \& Stella 1996).}. Furthermore, for both data-sets we performed the search in the 0.3--10\.keV energy band, and also dividing the entire data set in two energy bands (0.3--2 and
2--10\,keV).

For the \CXO\ data we performed an average over 7 FFTs with a bin
time of 2.85\,ms (see Fig.\,\ref{dps} right column), resulting in about 2,097,152 frequency bins for
each of the 7 averaged power spectrum\footnote{We could not use less FFTs because of computing limitations.}. The resulting power-spectrum had
a $\chi_{\nu}^{2}$ distribution with 14 degree of freedom (dof).

For the \XMM\ data we could perform a single FFT with a bin time of
60\,ms (over-sampling by a factor of 3 the timing resolution of the
instrument), resulting in 716,732 frequency bins in the power
spectrum. The resulting power-spectrum had a $\chi_{\nu}^{2}$
distribution with 2 dof (see Fig.\,\ref{dps} left column).

Note that given the long orbital period of \srcc\ (Bongiorno et al. 2011) with respect to the exposure time of the observations we report here, we do not need to de-modulate the photon arrival times for the orbital motion as we did for the 4\,day TeV binary \srcb\ (Rea et al. 2011).

In calculating the 3$\sigma$ detection upper limits reported in Fig.\,\ref{dps} we took into account for the number of bins searched, for the
different dofs of the noise power distribution, and for the
red-noise (Vaughan et al. 1994; Israel \& Stella 1996; Rea et al. 2010). We did not
find any periodic or quasi-periodic signal in any of the two observations. As a final try we attempted a joint search using both observations in
a single power spectrum, with again no detection of any signal. However, note that the $\sim$3\,years time-span between the two
observations would likely hamper the detection of a periodic signal
without the knowledge of its first derivative, and of the system precise orbital parameters.

We computed the 3$\sigma$ upper limits on the amplitude of a sinusoidal signal (which we define as pulsed fraction ($PF$)), according to Vaughan et al. (1994)
and Israel \& Stella (1996). These limits range in the 0.3--10\,keV energy band
between $PF<$22--34\% ($0.005-175$\,Hz) and 32--48\% ($0.005
-8$\,Hz), in the high and low X-ray state respectively (see
Fig.\,\ref{dps} bottom panels). We also infer the same limits as a function of the
energy band, which given the lower number of counts causes the
energy-dependent $PF$ limits to be slightly larger than those derived
using the whole energy range in the two datasets.

\subsection{Spectral analysis}
\label{spectral}

For the spectral analysis we binned both the \CXO\ and \XMM\ spectra such as to have at least 50 counts
per spectral bin (see Fig.\ref{spectra})\footnote{Note that the small difference in the derived spectral parameters we report in \S\ref{spectral} with respect to Hinton et al. (2009), is most probably due to a different spectral re-binning used in the analysis. Anyway, all the results of our analysis of the \XMM\ observation are consistent within a 99\% confidence level with those reported in Hinton et al. (2009). }.
We first fitted both spectra together with an absorbed power-law ({\tt phabs} and {\tt powerlaw} models under the {\tt XSPEC} version 12.5.0
spectral modeling program). All the absorption values we report here
are referred to abundances from Anders \& Grevesse (1989) and
photoelectric scattering cross-section from Balucinska-Church \&
McCammon (1998). Fitting the two spectra with the same model with all
parameters except the normalization value equal among the two spectra,
results in a $\chi_{\nu}^{2} = 1.07$ (190 dof; with $N_{\rm
  H} = (4.0\pm0.2)\times10^{21}$\,cm$^{-2}$, $\Gamma =
1.56\pm0.03$). Although the $\chi_{\nu}^{2}$ value is acceptable, the residuals of the \XMM\ observation are
clearly not satisfactory (see Fig.\ref{spectra}). This is mirror of the fact
that the large difference in counts among the two spectra make such a joint spectral fitting (and consequently the $\chi_{\nu}^{2}$ value) to be dominated by the \CXO\, data.

To solve the problem of having such bad residuals, we analize possible spectral variabilities between the two spectra. In particular, we re-fit the spectra keeping the absorption value ($N_{\rm H}$) equal among the two spectra and the power-law photon index ($\Gamma$) and
normalization free to vary. Although resulting in an acceptable
$\chi_{\nu}^{2} = 1.07$ (189 dof; with $N_{\rm H} =
(4.0\pm0.2)\times10^{21}$\,cm$^{-2}$, $\Gamma_{\rm cxo} = 1.57\pm0.03$
and $\Gamma_{\rm xmm} = 1.46\pm0.05$), the residuals of the \XMM\
observation were again not good (see Fig.\ref{spectra}). We then did a further trial
keeping equal the $\Gamma$ while leaving free the $N_{\rm H}$ and
normalization. Again we got an acceptable $\chi_{\nu}^{2} =1.08$ (189
dof; with $N_{\rm H, cxo} = (4.0\pm0.2)\times10^{21}$\,cm$^{-2}$,
$N_{\rm H, xmm} = (3.4\pm0.3)\times10^{21}$\,cm$^{-2}$, and $\Gamma =
1.56\pm0.03$), but bad residuals. 

Leaving all parameters free to vary for both observations we get a
good $\chi_{\nu}^{2} = 0.98$ (188 dof) and better, flat-looking residuals for both
\XMM\ and \CXO.
We then consider this as the best spectral modeling for both observations. In particular we 
find the following spectral parameters: for \CXO\ (during the high X-ray emission state), $N_{\rm H, cxo} =
(4.3\pm0.2)\times10^{21}$\,cm$^{-2}$, $\Gamma_{\rm cxo} =
1.61\pm0.03$, and an absorbed (unabsorbed) 0.3--10\.keV flux of
$(3.2\pm0.2)\times10^{-12}$\ergscm2 ($4.5\times10^{-13}$\ergscm2 ). For \XMM\ (during a low X-ray emission state): $N_{\rm H, xmm} =
(2.1\pm0.4)\times10^{21}$\,cm$^{-2}$, $\Gamma_{\rm xmm} =
1.18\pm0.08$, and an absorbed (unabsorbed) 0.3--10\.keV flux of
$(5.1\pm0.9)\times10^{-13}$\ergscm2 ($5.7\times10^{-13}$\ergscm2 ).

\begin{figure}
\hspace{-1cm}
\includegraphics[height=9cm,width=10cm]{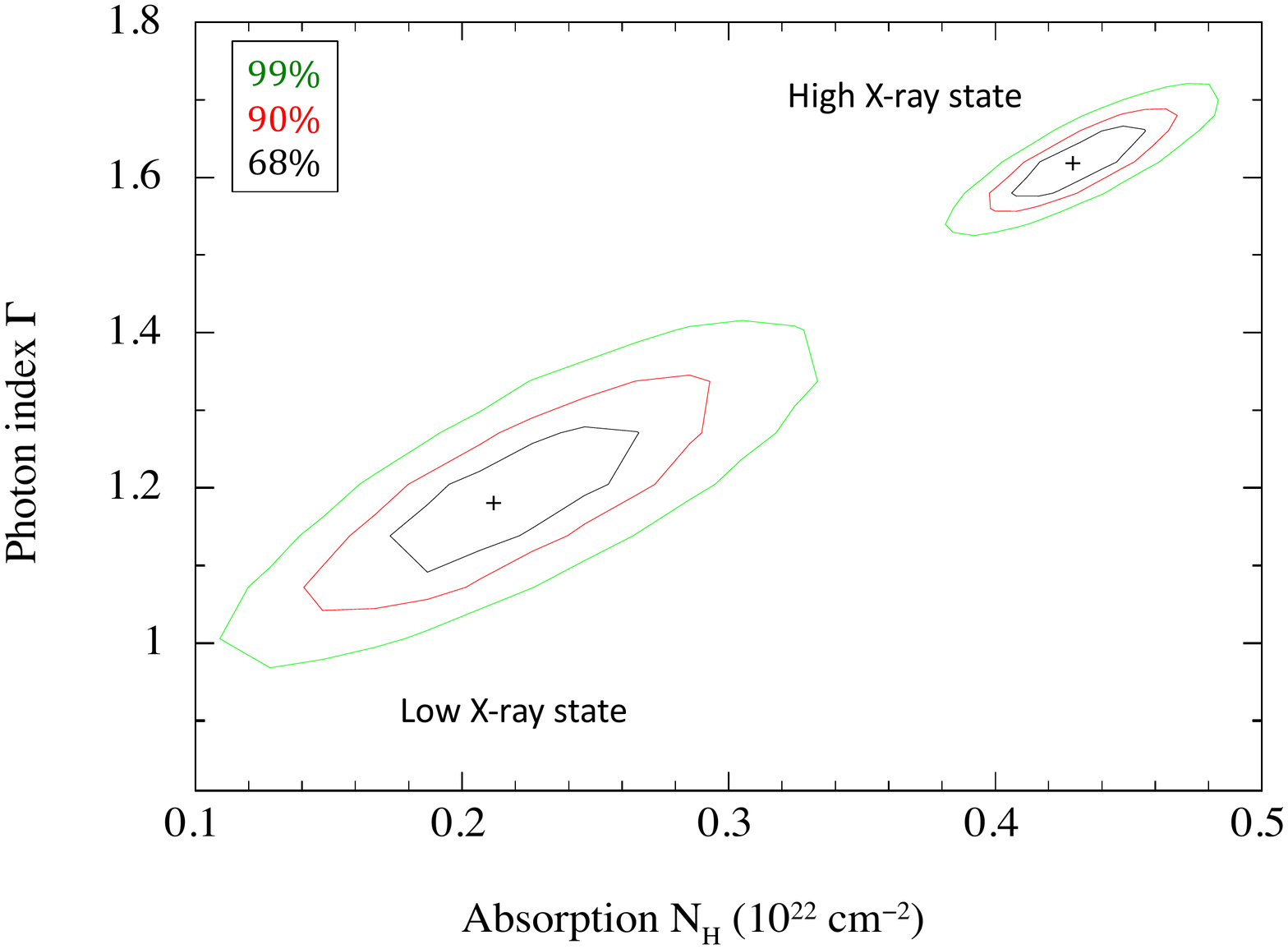}
\caption{Contour plots of the absorption value ($N_{\rm H}$) and photon index ($\Gamma$) derived from the fit of the \XMM\ spectrum taken while \srcc\ was in a low X-ray emitting state, and of the \CXO\ spectrum during a high X-ray state. Note that the $N_{\rm H}$ refers to abundances from Anders \& Grevesse (1989) and
photoelectric scattering cross-section from Balucinska-Church \& McCammon (1998).}
\label{contours}
\end{figure}

To investigate further the spectral change between the two observations, we plot the statistical contours for the absorption value and the power-law photon index (see Fig.\ref{contours}), which clearly shows that the spectral variability among the two X-ray emission state is significant at $>99$\% confidence level.

Assuming a distance of 1.5\,kpc (Hinton et al. 2009), the system had a luminosity change between the low and the high X-ray emission state of about an order of magnitude, with L$_{\rm low}\sim 1.54\times10^{32}$\ergs and L$_{\rm high}\sim 1.21\times10^{33}$\ergs .

\section{Concluding remarks}

We reported on a \CXO\ observation during the 2011 February increase of X-ray/TeV emission of the new TeV binary \srcc, which allowed us to perform the first detailed X-ray timing and spectral analysis of this source during its high state. As a comparison, we also studied the best X-ray data available in the archive for \srcc, taken while the source was in its low X-ray emission state (previously published in Hinton et al. 2009). We do not find any periodic or quasi-periodic signal from this system in any of those emission states, deriving a 3$\sigma$ upper limit on the X-ray pulsed fraction of \srcc\ of $\sim$30\% (highly dependent on the frequency, energy range and emission state; see \S\ref{timing} and Fig.\ref{dps} for details). The limits we derived for the X-ray pulsed fraction of \srcc\ are similar to those derived for the archetypical TeV binaries: \srca\ (Rea et al. 2010) and \srcb\ (Rea et al. 2011). Note that also the only firmly established TeV binary containing a pulsar,  PSR\,B1259--63, does not show X-ray pulsations (Chernyakova et al. 2009). This result shows that in the pulsar scenario, the pulsar emission itself cannot be the main responsible for the X-ray emission of these TeV binaries, which is instead likely dominated by the wind-wind or intra-wind shock (unless also in this case the putative pulsar is never pointing to us during its rotation).

Also for \srcc, we find here that there exists a significant spectral variability among the low and high X-ray emission states (the flux of the new \CXO\, observation is compatible with that observed by {\em Swift} in the same period; Bongiorno et al.  (2011)). In particular, comparing the two best available spectra, which were taken at different orbital phases, we can see $>3\sigma$ variability in the source spectral parameters. Furthermore, while $\Gamma$ increases from the low to high state by $\sim40$\% (hence the spectrum softens accordingly), the $N_{\rm H}$ increases by more than a factor of 2. Note also that for \srcc\ we find that the higher the flux, the steeper the X-ray  spectrum, which is at variance with the cases of \srcb\ and \srca, where the opposite behavior is found (Kishishita et al. 2009; Rea et al. 2010). In comparison with the other TeV binaries, we are tempt anyway to relate this spectral change with the orbital phase of \srcc. However our current ignorance on the orbital parameters of this system (beside the 320\,days period; Bongiorno et al. 2011) hamper for the moment any classification of the low and high states in terms of periastron, apastron, or any of the conjunctions.

\acknowledgments
This research has made use of data from the \CXO\ X-ray Observatory and software provided by the Chandra X-ray Center. NR is supported by a Ramon y Cajal Research Fellowship to CSIC.  We acknowledge the \CXO\ Director, Dr. H. D. Tananbaum, for granting us Director's Discretionary Time that allowed for this research to be done. We thank G.L. Israel for allowing the use of his {\tt DPS} software and useful discussion, D. Hadasch and A. Caliandro for a quick visibility check when the source was discovered at a high flux, and the anonymous referee for useful suggestions. This work has been supported by grants AYA2009-07391 and SGR2009-811, as well as the Formosa Program
TW2010005.

\label{lastpage}

\end{document}